\documentclass[pra,nofootinbib,notitlepage]{revtex4-1}
\pdfoutput=1
\usepackage{amssymb}
\usepackage{amsmath}
\usepackage{amsfonts}
\usepackage{amsthm}
\usepackage[stable]{footmisc}
\usepackage{hyperref,moredefs,graphicx,enumitem,verbatim}
\usepackage{bm,latexsym,array,subfigure}
\usepackage{tikz}
\usetikzlibrary{arrows,fit,positioning,automata,fadings,arrows.meta}
\usetikzlibrary{decorations,decorations.pathmorphing,decorations.pathreplacing,shapes.arrows,shapes.geometric}

\usepackage[framemethod=TikZ]{mdframed}

\newenvironment{Definition}[1][Definition]{\begin{trivlist}
\item[\hskip \labelsep {\bfseries #1}]}{\end{trivlist}}

\let\baraccent=\= % rename builtin command \= to \baraccent
\renewcommand{\=}[1]{\stackrel{#1}{=}} % for putting numbers above =

\begin{document}

\title{A Formal Model for Adaptive Free Choice in Complex Systems}

\author{Ian T. Durham}
\email[]{idurham@anselm.edu}
\affiliation{Department of Physics, Saint Anselm College, Manchester, NH 03102}
\date{\today}

\begin{abstract}
In this article, I develop a formal model of free will for complex systems based on emergent properties and adaptive selection. The model is based on a process ontology in which a free choice is a singular process that takes a system from one macrostate to another. I quantify the model by introducing a formal measure of the `freedom' of a singular choice. The `free will' of a system, then, is emergent from the aggregate freedom of the choice processes carried out by the system. The focus in this model is on the actual choices themselves viewed in the context of processes. That is, the nature of the system making the choices is not considered. Nevertheless, my model does not necessarily conflict with models that are based on internal properties of the system. Rather it takes a behavioral approach by focusing on the externalities of the choice process.
\end{abstract}

\maketitle
\vspace{-12pt}
\section{Introduction}
What is it that we mean when we ask if a system possesses free will? In most discussions of free will, the nature of what freedom entails is often taken for granted; if we ask if a given choice is free, we are assuming that it is a well-formed question to begin with. Rather than asking if a given choice is free, we might instead ask what, in general, a free choice \textit{is}. In other words, a more formal and rigorous definition of freedom of choice ought to be a prerequisite to any deeper understanding of free will.

This is not merely academic. In EPR tests in quantum mechanics, it is often assumed that the experimenters have free will (see~\cite{Fine:2017aa} for an overview). On the other hand, as Bell suggested, one could replace the experimenters with a pair of machines capable of making suitably random measurements~\cite{Bell:2004aa}. The variables measured by the machine are the same as the variables measured by the experimenters, but do machines carrying out pre-programmed algorithms to produce random measurements count as having free will? While they might produce measurements that are provably more randomly chosen than those chosen by experimenters, it is hard to say if a presumably non-conscious entity can possess free will. Therein lies the problem. Does it even make sense to refer to the machine's actions, which are guided by a pre-programmed algorithm, as making a choice? As Nozick has rightly pointed out, ``[a]n action's being non-determined is not sufficient for it to be free--it might just be a random act''~\cite{Nozick:1995aa}. A random act is no more free than a fully determined one.

Thus, it is that freedom of choice has a bearing on the nature of consciousness. Does our conception of consciousness drive our definition of free will or is it the other way around? Explicating the relationship between consciousness and volition has moved well beyond the theoretical. Famously, the Libet experiments suggest to some that volition can be an unconscious act~\cite{Libet:1985aa}. Additional experiments have built on Libet's work~\cite{Soon:2008aa,Maoz:2014aa}. On the other hand, our judgment of what this means appears to be driven by the outcomes. Specifically, experimental work by Shepherd suggests that the conscious causation of behavior tends to be judged as being free even when the causation is explicitly deterministic~\cite{Shepherd:2012aa}. 

The question of whether free will is compatible or incompatible with (causal) determinism has long been central to the nature of free will, at least in the Western perspective~\cite{:2002aa}. However, the indeterminacy of quantum systems raises similar issues for seemingly opposite reasons~\cite{Conway:2006aa,Conway:2009aa}. In other words, one could argue that both causal determinism as well as quantum indeterminacy are incompatible with free will. On the one hand, one can deny the existence of free will on the grounds that all choices are pre-determined in some way. On the other hand, one can deny the existence of free will on the grounds that all choices are simply macro-manifestations of random quantum processes, i.e., the universe is nothing but a collection of randomly fluctuating quantum fields. However, this misses a deeper point captured succinctly by O'Connor when he notes that ``though freedom of will requires a baseline capacity of choice \ldots the \textit{freedom} this capacity makes possible is, nevertheless, a property that comes in \textit{degrees} and can vary over time within an individual''~\cite{OConnor:2009aa} p. 183 (original emphasis). The usual debate over the existence of free will is thus really a debate over the capacity of choice. Less attention is paid to the nature of the freedom that this capacity makes possible. It is this freedom that is the motivation for the model described in this article. 

In order to build a model of the freedom engendered by the existence of a capacity of choice, it is necessary to assume that such a capacity exists. As such, the model that follows sidesteps the question of whether or not free will actually exists. Rather, it proceeds from the assumption that capacity of choice exists for some systems and develops a measure for the freedom that follows from this assumption. The model is based on a process ontology in which a \textit{free choice} is a process that takes a system from one macrostate to another. I lay the groundwork for this by first describing the self-evident characteristics of what I will call \textit{adaptive} free choices. This very roughly corresponds to what O'Connor calls ``willing'' or the ``conscious forming of an intention to act'' (see~\cite{OConnor:2009aa}, p. 178) though it does not explicitly presume an agent or system is conscious. I then develop the process ontology on which the formal model is constructed where I show that both determinism and indeterminism have a role to play in the nature of free choices. The formal model is then quantified by the $\zeta$-function as a measure of free choice and the $Z$-function as a corresponding measure of free will which is taken to be an aggregation of free choices. Free will in this model, then, is not taken to be the capacity of choice, but rather is viewed as an aggregation of the freedoms that a capacity of choice, if it exists, would make possible. Finally, I discuss methods for assessing certain aspects of the model and for assigning values to certain variables within the model.

%%%%%%%%%%%%%%%%%%%%%%%%%%%%%%%%%%%%%%%%%%
\section{Adaptive Free Choice}\label{AFW}
What do we generally think of when we think of free will? O'Connor claims that systems that possess free will should generally be possessed of three capacities: (1) an awareness and sensitivity to reasons for actions; (2) an ability to weigh and critically probe desires and intentions, and possibly to reevaluate goals; and (3) the ability to choose, based on reasons, which action to take on a given occasion~\cite{OConnor:2009aa}. While O'Connor claims that those are necessary for a system to possess free will, are they sufficient?

Suppose that I open my refrigerator with the intent of having something to eat and am presented with a variety of options. It is immediately obvious that, if my refrigerator only contained one thing, then I wouldn't have much of a choice. It seems clear, then, that, aside from any internal capacities that I might possess, one very clear external factor in assessing the freedom of a choice is the number of options that are available to choose from. By its very nature, the word ``choice'' implies that there must be a minimum of two options to choose from. However, it is also necessary that I be able to read those options (or a certain subset of them) into my memory and then process them, all in a finite amount of time (it wouldn't really be much of a choice if it took me an infinite amount of time to read in all the options). As such, there is a computational aspect to free choice. Indeed, several computational models of free will exist~\cite{Berto:2017aa,Hadley:2018aa}. However, we also assume that, for free will to exist, the system or agent experiencing it must have enough time to assign meaning or a value judgment (i.e., a weight) to each possible choice. In other words, if the possible choices are read into the system's memory too quickly, one could easily attribute any subsequent choice to mere instinct. O'Connor identifies three types of ``will'' or ``desire'': (1) minimally voluntary action, (2) willing or the conscious formation of an intention to act (briefly mentioned above) and (3) an urge or want~\cite{OConnor:2009aa}. It seems clear, then, that the capacity to weigh and probe desires and intentions, i.e., to assign meaning or a value judgment to each possible choice, is linked to the distinction between a minimally voluntary action and a ``willing''. What I refer to as an \textit{adaptive} free choice, then, includes the capacity to weigh and probe desires, the formation of an intention to act based on the action of weighing and probing desires, and finally the carrying out of the action whose intention was formed. That is, I assume that it is not merely sufficient to possess the capacity of choice. Some action representing the choice being made must also take place. This is not a trivial point.

Returning to the refrigerator, for the sake of simplicity, let's suppose that I only have two options---carrots and peppers. With only two choices, there is no real concern about the reading and processing time being too lengthy and we can assume, for the moment, that neither is it too short, i.e., I am free to stand in front of the refrigerator and ponder these two possibilities, assigning meaning to them in the form of weights. Let us suppose that I then make a choice---carrots, for the sake of argument. There is an often overlooked assumption in discussions of free will having to do with the nature of actually \textit{enacting} a choice. That is, we assume (usually implicitly) that, to a generally high degree of probability, enacting the choice results in the desired outcome. To put it another way, in choosing carrots, I am relying on the fact that there is a very high probability that, at some point between choosing carrots and actually removing them from the refrigerator and ingesting them, they \textit{remain carrots} and do not spontaneously turn into peppers. That is, we rely on the fact that the universe is a relatively stable place. Consider what would happen if we could \textit{not} be confident that our choices lead to their expected outcomes the vast majority of the time. If this were the case, we would likely give up on choosing since the outcomes would be closer to random and thus the act of making a choice would be pointless. As Nozick pointed out, a random act is not a free act~\cite{Nozick:1995aa}. Thus, there must be some level of determinism involved in free choices in the sense that the chosen action itself is nearly deterministic. The crucial point, however, is that each possible choice \textit{is a different action}. Thus, it is that we are choosing between different \textit{processes} rather than labeling each choice as a different outcome of the \textit{same} process~\cite{Durham:2018aa}. In other words, in order for the choice to be truly free, we have to have a high degree of confidence that the state that we finally choose actually occurs. Otherwise, there is no point in making a choice in the first place. However, this means that, when presented with a set of possible choices, each choice must represent a fundamentally different process, e.g., the actual process of reaching into my refrigerator for a carrot is not the same as the actual process of reaching into my refrigerator for a pepper (at the very least, they are not in the exact same location and thus the processes involve different spatial coordinates).

These characteristics of free choices are partially behavioral since they are based on the system's actions and its reaction to those actions. We weigh our choices based partly on past experience. If I have eaten both carrots and peppers before, I will know what they taste like and can use that information to place weights on each option. The fact that previous behavior can inform future behavior is a characteristic of many downwardly causal processes. This makes the weighting of the choices and the existence of a memory key differentiators, and motivates the use of the term \textit{adaptive} (see~\cite{Ellis:2016aa}). This distinguishes them from O'Connor's minimally voluntary actions which I interpret as upwardly causal. A minimally voluntary action is distinguished from a purely instinctual one by the fact that it still leads to a desired outcome (see~\cite{OConnor:2009aa} for a fuller discussion).

If we expect the outcomes of chosen actions to be nearly deterministic in most instances, how can we reconcile this with the fact that the quantum fields of which the universe is constructed, display a certain level of indeterminacy? Is it possible to obtain deterministic outcomes from sets of random processes? This, in fact, is exactly what happens in the thermodynamic limit of statistical physics. I use similar methods to construct the process ontology that is at the heart of this model.

%%%%%%%%%%%%%%%%%%%%%%%%%%%%%%%%%%%%%%%%%%
\section{Process Ontology}
In order to reconcile a seemingly deterministic macroworld with a fundamentally random microworld, I introduce a process ontology in which we may classify choices. The fundamental entities of this ontology are \textit{systems}, \textit{states}, and \textit{processes}. However, I refer to this as a \textit{process} ontology in order to emphasize the fact that the aim is to model both the internal reasoning about possible actions as well as the fulfillment of a chosen action.

A system $\sigma$ in this ontology can be anything one chooses it to be. We define states as follows.
\begin{Definition}[State]
A state $\bm{\psi}(\lambda)$ is a unique vector configuration of the fundamental components of a system as specified by some variable or set of variables $\lambda$.
\end{Definition}\noindent
The states of systems are thus functions of variables $\lambda$ which allow us to specify the state. Processes, then, take states from one configuration to another by changing $\lambda$.
\begin{Definition}[Process]
A process $\pi$ is a means by which a system may transition between some fixed state $\bm{\psi_i}(\lambda_i)$ and one (and only one) of $n$ possible states $\bm{\psi_j}(\lambda_j)$.
\end{Definition}\noindent
That is, though a process always ends in a single definite state, it may probabilistically end up in one of a possible collection of states. We assume that the outcome state is definite (i.e., is not a superposition). A graphical representation of a process is shown in Figure~\ref{process}.
\begin{figure}[h]
\centering
\begin{tikzpicture}[>=stealth, thick]
\node at (-1.25,0) {$\pi_{ij}$:};
\node at (-0.25,0) {$\bm{\psi}_i(\lambda_i)$};
\draw[gray,->] (0.25,0) -- (1.75,1);
\node at (2.625,1) {$\bm{\psi}_{j=1}(\lambda_{j=1})$};
\draw[gray,->] (0.25,0) -- (1.75,0.5);
\node at (2.625,0.5) {$\vdots$};
\draw[->] (0.25,0) -- (1.75,0);
\node at (2.625,0) {$\bm{\psi}_{j =k}(\lambda_{j=k})$};
\draw[gray,->] (0.25,0) -- (1.75,-0.5);
\node at (2.625,-0.45) {$\vdots$};
\draw[gray,->] (0.25,0) -- (1.75,-1);
\node at (2.625,-1) {$\bm{\psi}_{j=n}(\lambda_{j=n})$};
\end{tikzpicture}
\caption{\label{process} A process $\pi_{ij}$ takes a system $\sigma$ from some fixed state $\bm{\psi_i}(\lambda_i)$ to one (and only one) of $n$ possible states $\bm{\psi_j}(\lambda_j)$.}
\end{figure}
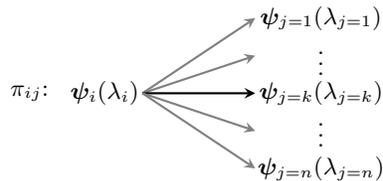\noindent
The \textit{multiplicity} $\omega (\pi_{ij})$ for process $\pi_{ij}$ is then defined to be the number of possible outcomes, i.e., $\omega (\pi_{ij}) = n$.

These definitions now allow us to describe a spectrum of processes. It might be true, for instance, that, for a given process $\pi_{ij}$, the multiplicity is $\omega (\pi_{ij}) = n = 1$. That is, there might be only one possible outcome for the process. Such a process is thus fully deterministic. Conversely, it may be true that $\omega (\pi_{ij}) = n > 1$ \textit{and} that all outcomes have an equal likelihood of occurring, i.e., the probability distribution for the process is flat. Such a process would naturally be fully random since no single outcome is favored over any other.

It is important to distinguish here the difference between determinism and causality. In this model, all processes are causal in that they follow from other processes. This includes ``spontaneous'' processes which are understood to follow from conditions produced by other processes. For example, spontaneous particle creation can only occur because of the presence of the underlying fields. This model would assume that the existence of those fields came about from some process, even if the nature of that process is not known. Thus, though this model is causal, it is not fully deterministic. The distinction between causality and determinism has been rigorously delineated by D'Ariano, Manessi, and Perinotti, and we refer the reader to~\cite{DAriano:2014aa} for details.

I note that all these definitions can be broken into two possible categories---micro or macro---depending on the level of the system under consideration. This has the usual meaning from statistics and statistical mechanics. Thus,  for instance, we can speak of a pair of (fair) dice as having 36 different microstates and 11 different macrostates. In this case, the microstates are the various configurations that produce a given macrostate. A pair of (fair) dice is a fairly simple system in which we assume the fundamental constituents (the dice) do not interact. In more complex systems, interactions may occur between the various parts of the system. It is not necessary for interactions to occur, however, in order to obtain a deterministic macrostate from a collection of equally likely microstates.

Consider a simple two-state system $\sigma$ in a state $i=1$. Since it is a two-state system, any potential process only has two possible outcomes: either the existing state evolves to itself or to the other state. Let us assume that each of these states is equally likely to occur in the long run. Examples of such systems include two-state paramagnets, fair coins (being flipped), etc. Thus, the process $\pi_{1j}$ for the evolution of a simple two-state system is

\begin{equation}
\begin{tikzpicture}[>=stealth, thick]
\node at (-1.25,0) {$\pi_{1j}$:};
\node at (-0.25,0) {$\bm{\psi}_1(\lambda_1)$};
\draw[->] (0.35,0) -- (1.75,0.5);
\node at (2.75,0.5) {$\bm{\psi}_{j=1}(\lambda_{j=1})$};
\draw[->] (0.35,0) -- (1.75,-0.5);
\node at (2.75,-0.5) {$\bm{\psi}_{j=2}(\lambda_{j=2})$};
\end{tikzpicture}
\end{equation}
The multiplicity for this process is $\omega(\pi_{1j}) = 2$. If we assume that each of the two possible outcomes is equally likely, the probability distribution is quite simple. 

Now, consider a more complex system that is composed of smaller sub-systems. The more complex system will be referred to as a macrosystem and its states and processes will be referred to as macrostates and macroprocesses. I will distinguish these from the component microsystems, microstates, and microprocesses by using capital letters. Thus,  for example, if $\pi_{ij}$ is a microprocess taking a microsystem $\sigma$ from some fixed microstate $\bm{\psi_i}(\lambda_i)$ to one (and only one) of $n$ possible microstates $\bm{\psi_j}(\lambda_j)$, then $\Pi_{IJ}$ is a macroprocess taking a macrosystem $\Sigma$ from some fixed macrostate $\bm{\Psi_I}(\Lambda_I)$ to one (and only one) of $N$ possible macrostates $\bm{\Psi_J}(\Lambda_J)$.

As an example, consider a macrosystem $\Sigma_{AB}$ that consists of two \textit{non}-interacting two-state microsystems $\sigma_A$ and $\sigma_B$ that is in a state $I=1$. The possible outcome states and the associated probability distribution are shown in Figure~\ref{smalltwostate}:
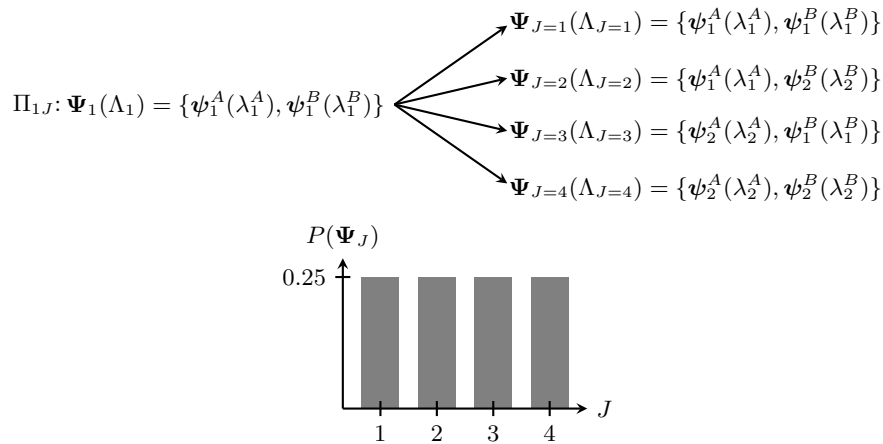
\begin{figure}[h]
\centering
\begin{tabular}{c}
\begin{tikzpicture}[>=stealth, thick]
\node at (-4.5,0) {$\Pi_{1J}$:};
\node at (-2,0) {$\bm{\Psi}_1(\Lambda_1)=\{\bm{\psi}^A_1(\lambda^A_1),\bm{\psi}^B_1(\lambda^B_1)\}$};
\draw[->] (0.25,0) -- (1.75,1.05);
\node at (4.25,1.1) {$\bm{\Psi}_{J=1}(\Lambda_{J=1})=\{\bm{\psi}^A_1(\lambda^A_1),\bm{\psi}^B_1(\lambda^B_1)\}$};
\draw[->] (0.25,0) -- (1.75,0.35);
\node at (4.25,0.35) {$\bm{\Psi}_{J=2}(\Lambda_{J=2})=\{\bm{\psi}^A_1(\lambda^A_1),\bm{\psi}^B_2(\lambda^B_2)\}$};
\draw[->] (0.25,0) -- (1.75,-0.35);
\node at (4.25,-0.35) {$\bm{\Psi}_{J=3}(\Lambda_{J=3})=\{\bm{\psi}^A_2(\lambda^A_2),\bm{\psi}^B_1(\lambda^B_1)\}$};
\draw[->] (0.25,0) -- (1.75,-1.05);
\node at (4.25,-1.1) {$\bm{\Psi}_{J=4}(\Lambda_{J=4})=\{\bm{\psi}^A_2(\lambda^A_2),\bm{\psi}^B_2(\lambda^B_2)\}$};
\end{tikzpicture}
\\
\begin{tikzpicture}[>=stealth, thick]
\fill[gray] (0.25,0) rectangle (0.75,1.75);
\fill[gray] (1,0) rectangle (1.5,1.75);
\fill[gray] (1.75,0) rectangle (2.25,1.75);
\fill[gray] (2.5,0) rectangle (3,1.75);
\draw[->] (0,0) -- (0,2);
\node[above] at (0,2) {$P(\bm{\Psi}_J)$};
\draw[->] (0,0) -- (3.25,0);
\node[right] at (3.25,0) {$J$};
\draw[-] (0.5,-0.1) -- (0.5,0.1);
\node[below] at (0.5,-0.1) {$1$};
\draw[-] (1.25,-0.1) -- (1.25,0.1);
\node[below] at (1.25,-0.1) {$2$};
\draw[-] (2,-0.1) -- (2,0.1);
\node[below] at (2,-0.1) {$3$};
\draw[-] (2.75,-0.1) -- (2.75,0.1);
\node[below] at (2.75,-0.1) {$4$};
\draw[-] (-0.1,1.75) -- (0.1,1.75);
\node[left] at (-0.1,1.75) {$0.25$};
\end{tikzpicture}
\end{tabular}
\caption{\label{smalltwostate} The macroprocess $\Pi_{1J}$ has four possible outcomes determined by the possible outcomes of each of the independent, non-interacting two-state subsystems. The probability of each possible outcome is the same. Recall that, once enacted, a process ultimately only leads to a single outcome, e.g., while the act of flipping a coin can lead to two possible states, once the flip has concluded, the coin ultimately ends up in only one state.}
\end{figure}
The probability of any given macrostate is given by the multiplicity of that macrostate divided by the total multiplicity. In this situation, there is only one microstate for each macrostate and so

\begin{equation}
P(\bm{\Psi}) = \frac{\Omega_{1(J=K)}}{\Omega_{1J}} = \frac{\Omega_{1(J=K)}}{\omega^A_{1j}\cdot\omega^B_{1j}} = \frac{1}{2\cdot 2} = 0.25
\end{equation}
where $\Omega_{1(J=K)}$ is the number of microstates that correspond to macrostate $J=K$ and where $\Omega_{IJ}=\omega^A_{1j}\cdot\omega^B_{1j}$ follows from the fact that multiplicities are multiplicative. Once again, the probability distribution is flat.

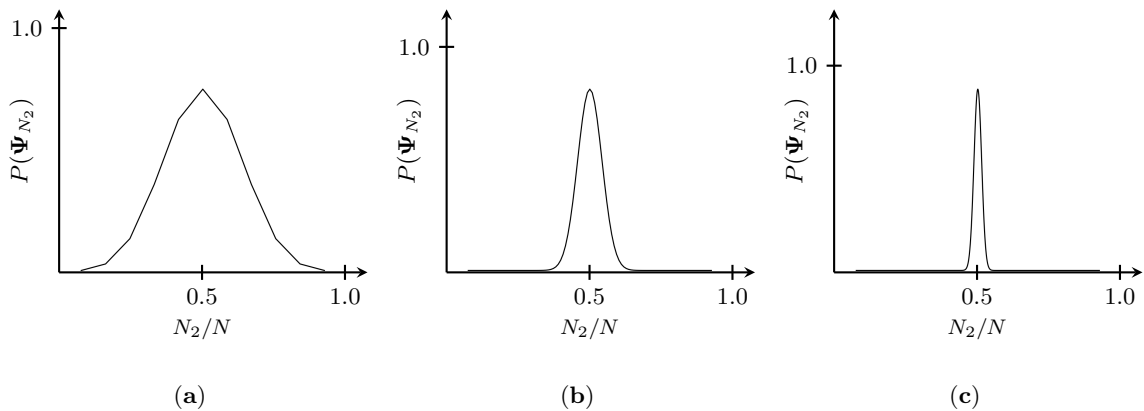
\begin{figure}[h]
\centering
\begin{tabular}{ccc}
\begin{tikzpicture}[>=stealth, thick]
\node at (1.85,1.25) {\includegraphics[width=1.8in]{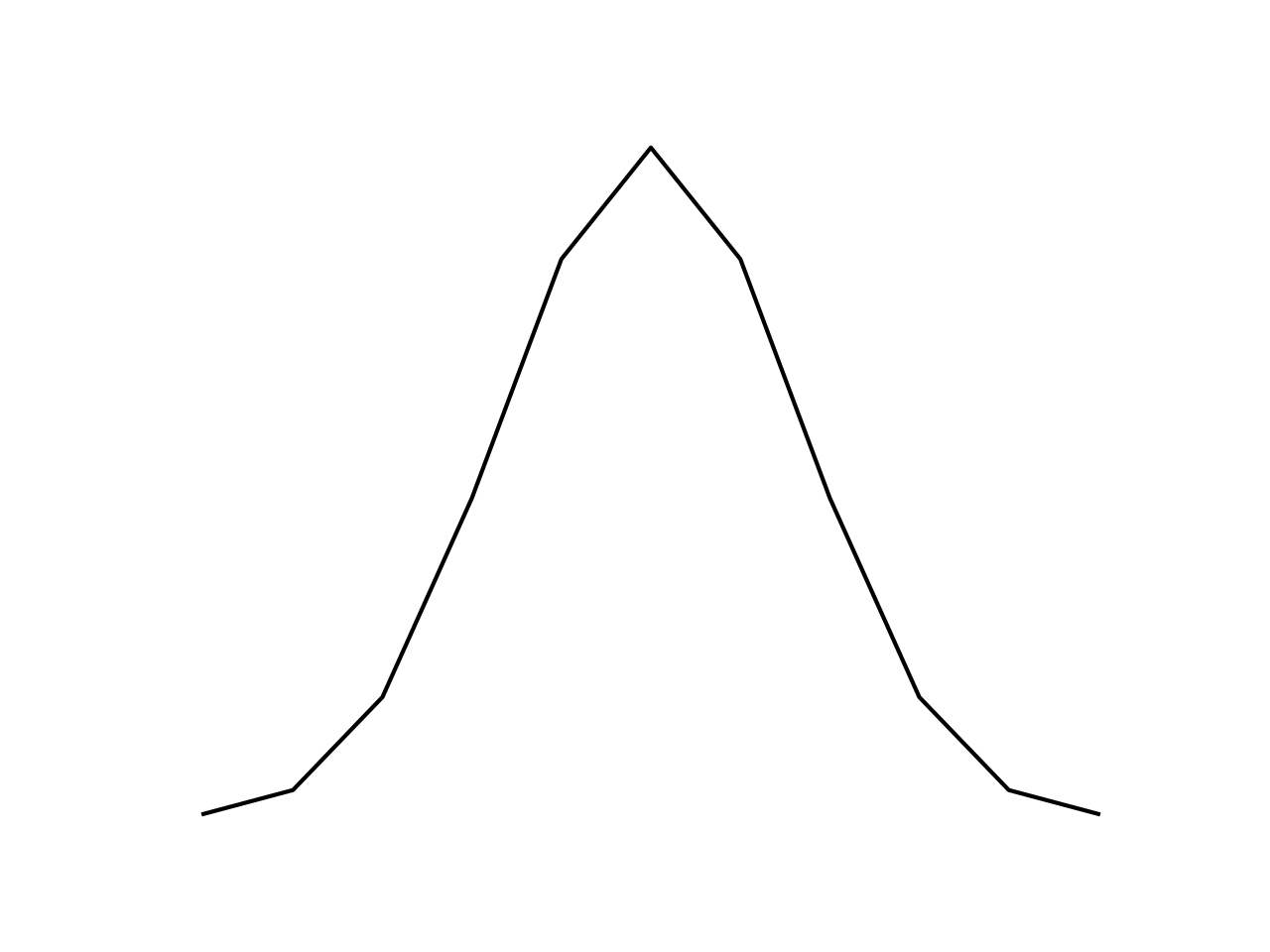}};
\node[rotate=90] at (-0.5,1.75) {$P(\bm{\Psi}_{N_2})$};
\node at (1.9,-0.75) {\footnotesize{$N_2/N$}};
\draw[->] (0,0) -- (4.1,0);
\draw[->] (0,0) -- (0,3.5);
\draw (1.9,-0.1) -- (1.9,0.1);
\node[below] at (1.9,-0.1) {0.5};
\draw (3.8,-0.1) -- (3.8,0.1);
\node[below] at (3.8,-0.1) {1.0};
\draw (-0.1,3.25) -- (0.1,3.25);
\node[left] at (-0.1,3.25) {1.0};
\end{tikzpicture}
&
\begin{tikzpicture}[>=stealth, thick]
\node at (1.85,1.25) {\includegraphics[width=1.8in]{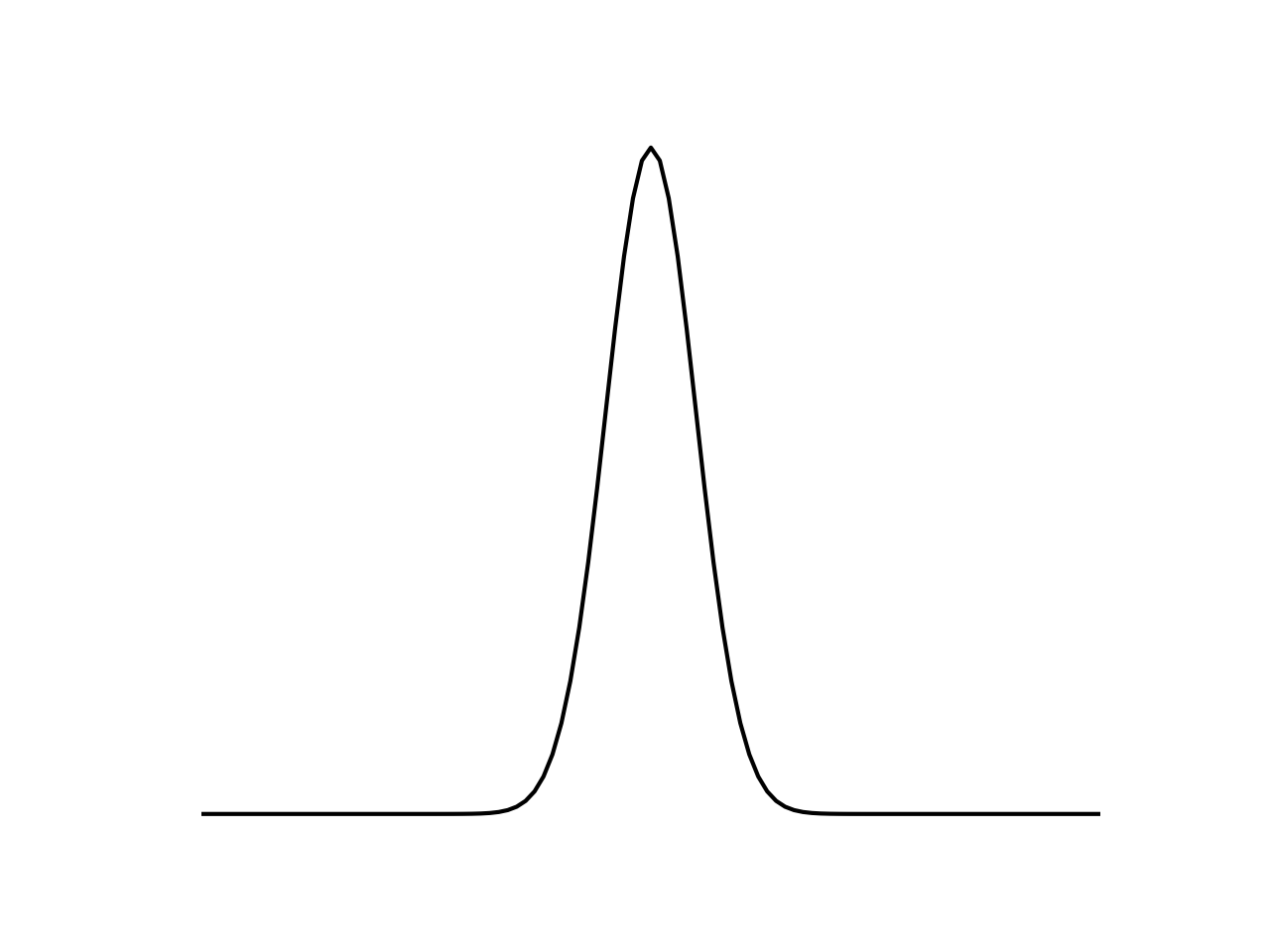}};
\node[rotate=90] at (-0.5,1.75) {$P(\bm{\Psi}_{N_2})$};
\node at (1.9,-0.75) {\footnotesize{$N_2/N$}};
\draw[->] (0,0) -- (4.1,0);
\draw[->] (0,0) -- (0,3.5);
\draw (1.9,-0.1) -- (1.9,0.1);
\node[below] at (1.9,-0.1) {0.5};
\draw (3.8,-0.1) -- (3.8,0.1);
\node[below] at (3.8,-0.1) {1.0};
\draw (-0.1,3) -- (0.1,3);
\node[left] at (-0.1,3.0) {1.0};
\end{tikzpicture}
&
\begin{tikzpicture}[>=stealth, thick]
\node at (1.85,1.25) {\includegraphics[width=1.8in]{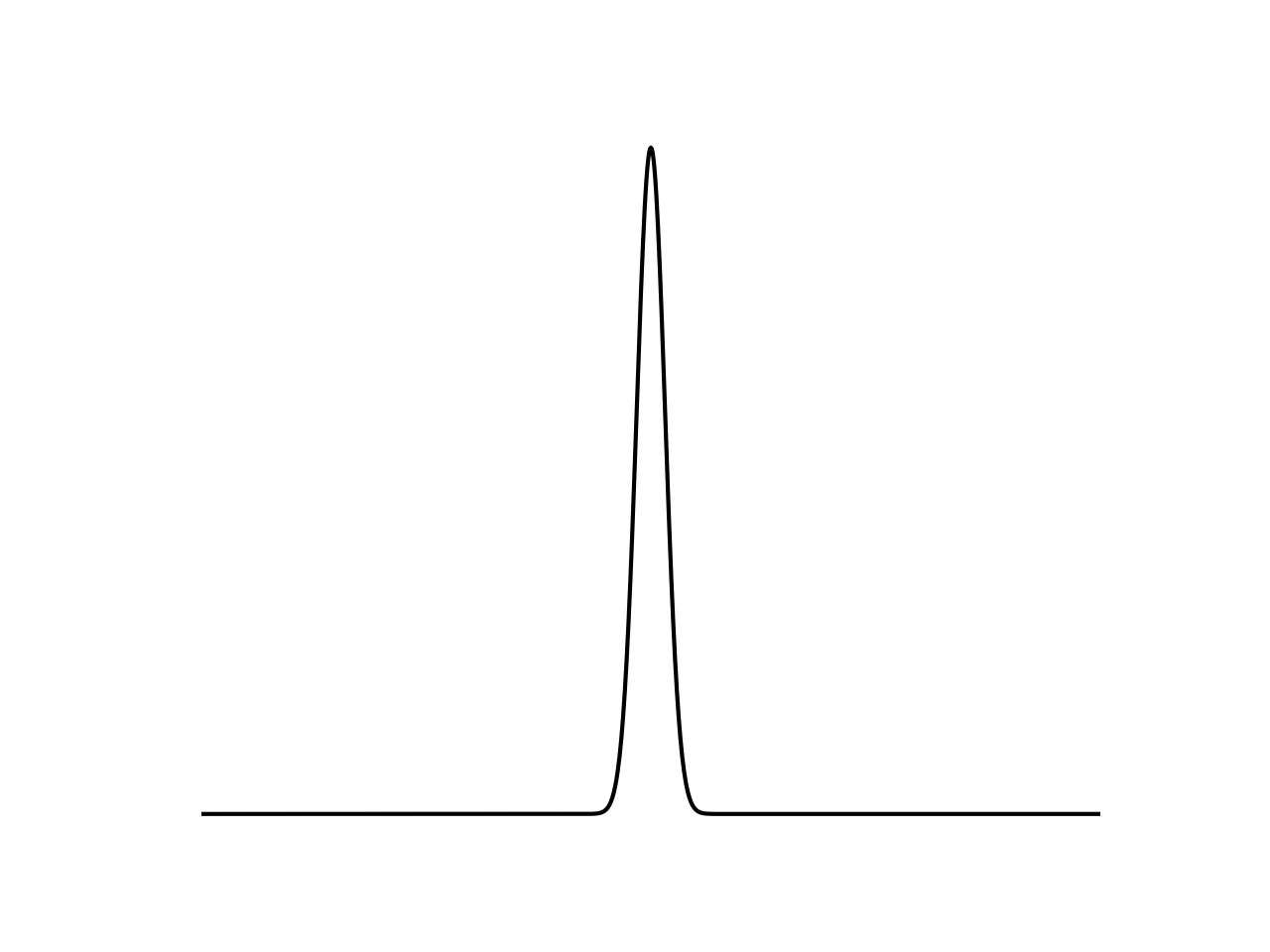}};
\node[rotate=90] at (-0.5,1.75) {$P(\bm{\Psi}_{N_2})$};
\node at (1.9,-0.75) {\footnotesize{$N_2/N$}};
\draw[->] (0,0) -- (4.1,0);
\draw[->] (0,0) -- (0,3.5);
\draw (1.9,-0.1) -- (1.9,0.1);
\node[below] at (1.9,-0.1) {0.5};
\draw (3.8,-0.1) -- (3.8,0.1);
\node[below] at (3.8,-0.1) {1.0};
\draw (-0.1,2.75) -- (0.1,2.75);
\node[left] at (-0.1,2.75) {1.0};
\end{tikzpicture}\\
& & \\
(\textbf{a}) & (\textbf{b}) & (\textbf{c})
\end{tabular}
\caption{\label{twostatemultplot} The normalized probability distribution for a two-state system is shown for (\textbf{a}) $N=10$, (\textbf{b}) $N=100$, and (\textbf{c}) $N=1000$ subsystems as a function of $N_2/N$. As can be seen, for larger values of $N$, a smaller number of macrostates tend to dominate over the rest. As $N\to\infty$, the distribution tends toward a delta function.}
\end{figure}

Now, consider an even more complex macrosystem composed of $N$ non-interacting two-state microsystems and let us ask how many of the subsystems will be in the state $j=2$ after the process $\Pi_{1J}$ occurs. The answer is given by the multiplicity, which is

\begin{equation}
\Omega_{1J} = \left(
\begin{array}{c}
N \\
N_{j=2}
\end{array}
\right) = \frac{N!}{N_{j=2}!(N-N_{j=2})!}.
\end{equation}
If $N$ is very large, we can use Stirling's approximation, $N!\approx N^Ne^{-N}\sqrt{2\pi N}$, which gives

\begin{equation}
\Omega_{1J} = \frac{N^N}{N_{2}^{N_2}(N-N_{2})^{(N-N_2)}}
\label{twostatemult}
\end{equation}
where $N_2 = N_{j=2}$. The probability $P(\bm{\Psi}_{N_2})$ of a state $\bm{\Psi}_{N_2}$ occurring is plotted as a function of $N_2/N$ for three different values of $N$ in Figure~\ref{twostatemultplot} where I have normalized each distribution (see~\cite{Schroeder:2000aa} Ch. 2).
We can see from this that, as the size of the system increases, despite the underlying processes being entirely random, the system tends to cluster in a small number of macrostates. As $N\to\infty$, the system tends toward a single macrostate. That is, the transition between state 1 and some state $J$ \textit{becomes deterministic}. From this we can see that sets of random microprocesses can lead to deterministic macroprocesses.

The processes considered in the above example were non-interacting. The real world, of course, includes interactions. An interaction between two systems is also a process that in some way changes the variables of each system. As such, each subsystem is undergoing a process that is influenced by the other subsystems' processes. The details of the process ontology for such situations is beyond the scope of the current work, which is intended as an introduction to the model. However, subsequent works will elaborate on the underlying process ontology and discuss how it affects the distributions. For the purposes of explicating the basic model of free will, the nature of the actual distributions is not important. The distributions become important when assessing the actual choices that are fed into the model.

%%%%%%%%%%%%%%%%%%%%%%%%%%%%%%%%%%%%%%%%%%

\section{The Model}
I will refer to any system that possesses the capacity of choice, as introduced in Section~\ref{AFW}, as an `agent'. A choice can only be said to be free if the agent can make some judgment about all possible choices in order to weigh them against one another, i.e., they must mean something to the agent. Otherwise, the choice is random (and thus meaningless). In addition, the agent has to have a high degree of confidence that the choice they finally make will actually be realized. Crucially, different choices represent different processes. I thus \textit{define} choice in the following manner.
\begin{Definition}[Choice]
A choice is a (possible) macroprocess $\Pi_{IJ}$ that is a means by which a system may transition between some fixed state $I$ and one (and only one) of $n$ possible states $J$ .
\end{Definition}\noindent
Note that the lower-case $n$ is deliberate and the reason will become evident in a moment.

For a particular choice to be free, the total number of choices presented to the agent at a given time, which I will call the `choice ensemble' $\Gamma$, must be read into the system's memory and then processed in a finite amount of time. Otherwise, the possibility of making  a choice is undecidable. Likewise, stability is assumed, as described in Section~\ref{AFW}, in that the outcomes of the actions associated with each choice do not change subsequent to the decision, i.e., carrots don't spontaneously turn into peppers.

Given the above definition, I add the following axioms:
\begin{itemize}[leftmargin=*,labelsep=5.5mm]
\item
\textbf{Axiom 1}.The most fundamental systems are irreducible to other systems, i.e., they contain no interactions and cannot be partitioned.
\item
\textbf{Axiom 2}. All possible configurations, i.e., microstates, of fundamental systems are equally likely in the long run.
\item
\textbf{Axiom 3}. A system's macrostates are formed via interacting (micro)processes.
\item
\textbf{Axiom 4}. The probability that a choice will lead to its macrostate is arbitrarily high if the choice is free, i.e., a free choice inevitably leads to a nearly deterministic outcome.
\item
\textbf{Axiom 5}. The number $N$ of possible choices that a system has must be small enough to be read into the system's memory in a finite time.
\item
\textbf{Axiom 6}. The choices do not change at any time during the agent's processing of the choices nor during the agent's enactment of the choice.
\item
\textbf{Axiom 7}. The choices must be distinct.
\end{itemize}
These are the \textit{formal} axioms of the model. There are certain assumptions and definitions that fully describe the model that will be introduced in the next section. However, from the formal standpoint, we restrict ourselves to these axioms. It is worth noting that I do \textit{not} require that the interactions of Axiom 3 be classical. This is intentional. It is entirely possible that such an assumption will lead to unusual results that require a refinement of the model, but that is an area for future work.

Now, suppose that some macrosystem $\Sigma$ composed of $k$ microsystems $\sigma_1,\ldots,\sigma_k$ is presented with an ensemble $\Gamma$ of $N$ possible choices $\Pi_{11},\ldots,\Pi_{1N}$. Each choice in the ensemble has $n_k$ possible outcomes (hence the choice of a lower-case $n$ in Definition 3). The probability distributions associated with the choices in the ensemble are $P_1(\bm{\Psi};\textrm{\textbf{E}}_1,\bm{\textrm{K}}_1),\ldots,P_N(\bm{\Psi};\textrm{\textbf{E}}_N,\textrm{K}_N)$ where $\textrm{\textbf{E}}_i$ is the mean and $\textrm{K}_i$ is the variance for the given distribution. The ensemble of choices may be represented as a mixed distribution,

\begin{equation}
\Gamma(\bm{\Psi}) = w_1P_1(\bm{\Psi};\textrm{\textbf{E}}_1,K_1) + w_2P_2(\bm{\Psi};\textrm{\textbf{E}}_2,K_2) + \cdots + w_NP_N(\bm{\Psi};\textrm{\textbf{E}}_N,K_N)
\label{ensemble}
\end{equation}
with weights $w_i \ge 0, \; w_1 + w_2 + \cdots + w_N = 1$ and where $\Gamma(\bm{\Psi})$ is a convex function.

A particularly useful analysis of multimodal distributions is based on the distribution's overall topology. Ray and Lindsay give a method that utilizes what they refer to as the ``ridgeline'' function which describes the shape of $\Gamma(\bm{\Psi})$ as

\begin{equation}
\bm{\Psi}^*(a) = \left[\sum_{i=1}^n a_i\textrm{\textbf{K}}_i^{-1}\right]\times\left[\sum_{i=1}^n a_i\textrm{\textbf{K}}_i^{-1}\textrm{\textbf{E}}_i\right]
\label{ridgeline}
\end{equation}
where $a$ belongs to the $n-1$ dimensional unit simplex $\mathcal{S}_n=\{a\in\mathbb{R}^n:a_i\in[0,1],\sum_{i=1}^n a_i = 1\}$. Here, $\textrm{\textbf{K}}_i\in\mathbb{R}^{D\times D}$, $\textrm{\textbf{E}}_i\in\mathbb{R}^D$ for a $D$-dimensional space and $\textrm{\textbf{K}}_i$ and $\textrm{\textbf{E}}_i$ correspond to the covariance and mean of the $i$th component, respectively. This function can be used to identify the locations of the peaks in the distributions if they are not clearly known~\cite{Ray:2005aa}. Here, I will assume that the peaks are reasonably distinct in that their individual means and variances, and thus their locations in the overall distribution, are known.

Axiom 7 above captures the assumption that choices can only be said to be free if they are reasonably distinct. If an agent is presented with two essentially identical choices, then there are fewer criteria by which the agent can distinguish them and thus their distributions will have a certain amount of overlap. If spatial information is included in their distributions, then they won't overlap perfectly since the two choices will, at the very least, represent different spatial locations. For example, in choosing between two identical carrots, one is choosing between two objects that, though perhaps otherwise identical, are in different spatial locations. In any case, one expects that the more distinct the choices, the more free the ability to choose; if the choices are identical, the ability to choose is closer to a random process, e.g., the choice between two nearly identical carrots is not as free as the choice between a carrot and a pepper. A convenient measure of the distinctness of the choices in this case is given by the Mahalanobis distance~\cite{Mahalanobis:1936aa,Ray:2005aa} between any pair of constituent distributions $i$ and $j$ in the mixed distribution $\Gamma(\bm{\Psi})$ given by

\begin{equation}
\Delta_M(\textrm{\textbf{E}}_i,\textrm{\textbf{E}}_j,\textrm{\textbf{K}}) = \sqrt{(\textrm{\textbf{E}}_j-\textrm{\textbf{E}}_i)^{T}\textrm{\textbf{K}}^{-1}(\textrm{\textbf{E}}_j - \textrm{\textbf{E}}_i)}
\label{maha}
\end{equation}
where $\textrm{\textbf{E}}_i$ and $\textrm{\textbf{E}}_j$ are the respective means and $\textrm{\textbf{K}}$ is the covariance matrix. When the covariance matrix is diagonal, this reduces to the standardized Euclidean distance. The Mahalanobis distance is preserved under full-rank linear transformations of the space that is spanned by the data comprising the distributions. Intuitively, then, the larger the value of $\Delta_M$, the more distinct choices $i$ and $j$ will be. Within the full ensemble of choices $\Gamma(\bm{\Psi})$, the freedom of choice $i$ depends on the \textit{minimum} distance $\Delta_M$ between $i$ and each other choice in the ensemble $\min (\Delta_M)_i$.

Capturing Axiom 5 proves to be a bit tricky. We could quite simply define a time function $T$ that depends on the total number of choices $N$ and require that it be finite. However, one might suppose that different values for $\Delta_M$ or even $\bm{\Sigma}_i$ will affect the time required to read in and process the choices. For instance, if two neighboring choices are not particularly distinct in that they have considerable overlap in their individual distributions, this might lead to the agent spending more time processing and assigning weights to these two choices. Thus, I will assume that the time it takes for the agent to read in and process the choices is some function of their number $N$ as well as the general topology $\bm{\Psi}^*$ of the distribution, i.e., $T(\bm{\Psi}^*,N)$. The crucial requirement is that 

\begin{equation}
t_{min} < T(\bm{\Psi}^*,N) < \infty
\label{time}
\end{equation}
where $t_{min}$ is some minimum time below which the choice becomes either minimally voluntary or purely instinctual. This ensures that the information can be read into memory and processed in a finite amount of time. 

It is debatable whether, aside from requiring that $T$ be finite, we also need to require that it be minimized but greater than $t_{min}$. Obviously if $T$ is too small, the system is not extracting meaning from the process and thus the choice is minimally voluntary or instinctual. Hence the lower bound on $T$. On the other hand, one could argue that a shorter time means the system is more efficient. Greater efficiency does not necessarily mean less freedom. Thus it is that we also might require that $T$ be as close to $t_{min}$ as possible without dropping below it. I will leave this point unanswered in this initial version of the model. However, I discuss some related points in Section~\ref{choice}.

The covariance, time function, and minimum Mahalanobis distance, then, provide a means of measuring Axioms 4, 5, and 7, respectively. This leads us naturally to a measure of the ``freedom'' of any particular choice in the ensemble.
\begin{Definition}[Free choice]
Given a finite number of possible choices $N$ from ensemble $\Gamma(\bm{\Psi})$, the ``freedom'' of choice $i$ is given by the function

\begin{equation}
\zeta(\textrm{\textbf{K}}_i,T,\min (\Delta_M)_i)_i = C_i \times \textrm{\textbf{K}}_i^{-1}\times T(\bm{\Psi}^*,N)\times \min (\Delta_M)_i
\label{freechoice}
\end{equation}
where $C_i$ is a constant of proportionality with units of inverse time. 
\end{Definition}\noindent
I have assumed that maximizing the freedom of the choice entails minimizing its covariance since this increases the ``sharpness'' of the probability distribution within the ensemble. The larger the value of the $\zeta$-function, then, the more free the choice.

This, of course, is only a measure of the freedom of a single choice. It is entirely possible that an agent could have many free choices and many non-free choices. One would assume that a majority of choices for any agent said to possess \textit{free will} would have a high measure of freedom. The question is whether freedom in this sense is additive or multiplicative. We can answer this by returning to the statistical arguments that underpin the first few axioms.

In a certain sense, the freedom of a given choice is related to how easy it is to actually make that choice. Roughly speaking, then, the more ways in which one can make a specific choice, the greater the freedom of that choice. In that sense, freedom is similar to the outcomes of a process. The number of outcomes for a given macrostate of a process is measured by the multiplicity which is multiplicative. Thus, we can, by analogy, take freedom to be multiplicative in this particular model. As such, we can take a measure of \textit{free will} to be the following.
\begin{Definition}[Free will]
Given some number of processes that result in choices, an agent's free will is given by a partition function

\begin{equation}
Z(\zeta) = \prod_{i=1}^{N}\zeta(\textrm{\textbf{K}}_i,T,\min (\Delta_M)_i)_i,
\label{freewill}
\end{equation}
i.e., the level of free will that a system possesses over $N$ choices is dependent in a multiplicative way on the level of freedom in each of those choices.
\end{Definition}

\section{Weighting the Choices}\label{choice}
A crucial component of the model is the set of weights given in Equation~\eqref{ensemble}. It is within these weights that the action of choosing can either be entirely random (if the weights are all equal), entirely deterministic (if there is only a single constituent distribution), or something in between. Freedom is a spectrum between entirely random and entirely deterministic and it is this spectrum that $\zeta$ and $Z$ measure.

There are two basic methods for assigning the weights in the function $\Gamma(\bm{\Psi})$ from a physical standpoint. In the one case, the weights can be derived from the \textit{internal} dynamics of the agent. That is, we could begin with a set of $k$ fundamental subsystems each with some state $\bm{\psi}(\lambda)_i$. These subsystems could be allowed to evolve naturally through various processes $\pi_{ij}$ such that the agent itself evolves by some macroprocess $\Pi _{IJ}$ between two macrostates $\bm{\Psi}_I$ and $\bm{\Psi}_J$, where lowercase labels represent microstates and uppercase labels represent macrostates. It is entirely possible that the agent \textit{could} have evolved via an entirely different macroprocess $\Pi _{IJ^{\prime}}$ to macrostate $\bm{\Psi}_{J^{\prime}}$ under the same internal conditions since the outcomes of the individual microprocesses are assumed to be random in accordance with Axiom 2. Each of these possible macrostates may have an associated probability distribution. Over a large number of evolutions of the agent to these macrostates, the weights can be established statistically and an ensemble $\Gamma(\bm{\Psi})$ can be formed. 

Establishing the weights in this instance relies solely on the internal dynamics of the agent and is an entirely upwardly causal process. It would be easy to dismiss this as a free choice either on the grounds that, as a closed system, the agent is causally deterministic or on the grounds that it is simply reacting to underlying random processes. It is important to note that this does not mean the system is free from external interactions. It simply means that the internal dynamics are what drives the state evolution. The external constraints may be what sets the initial conditions in the form of the initial state $\bm{\Psi}_I$ but plays no role in the determination of the actual weights.

However, no system is truly closed and so it is more accurate to assume that environmental and contextual factors also drive the evolution of the processes that the agent undergoes. In actuality, the agent interacts with its environment \textit{during} the macroprocesses such that the dynamics are influenced by the environment in \textit{addition} to internal factors. Axiom 2 is still satisfied since the underlying microstates of both the agent and its environment remain equally likely in the long run, but conditions are now contextual such that the \textit{macrostates} are now partially constrained by external factors. These external factors can be construed as assigning a form of \textit{meaning} to the weights. This is a form of downward causation as discussed by Ellis and others (see, for example,~\cite{:2009aa}). Specifically, this can be viewed as a form of decoupling similar to dynamic or symbolic decoupling that has been argued as a possible driver for free will~\cite{Juarrero:2009aa}. However, there is still something missing.

Let's return to the choice between carrots and peppers in my refrigerator. If I choose carrots, did I do so because my genetics and specific worldline in spacetime predisposed me to do so? If so, it was hardly a free choice. Indeed, this is a typical feature of arguments \textit{against} free will (see, for example,~\cite{Newsome:2009aa}). On the other hand, perhaps I chose the carrot because I consciously weighed my past experience against how I ``felt'' at the time. The difference between the two is that the second involves accessing \textit{memory} in the broad sense (i.e., including both stored practical information as well as potentially qualia). One might argue that both instances involve memory, but that really isn't the case. Accessing memory is a necessarily downwardly causal process whereby a system in a higher-level state actively accesses stored information about lower-level states. Specifically, the process of choosing from a set of choices, which necessarily includes assigning them weights and accessing memory, is precisely a form of adaptive selection~\cite{Holland:1992aa}. Adaptive selection involves a process of variation (the process ontology in this model) that generates an ensemble of states (here these are the macrostates) from which an outcome is selected according to some kind of selection criteria. Indeed, the selection criteria themselves can be chosen through a form of adaptive selection~\cite{Ellis:2016aa}. The presence of memory is fundamental to any adaptive process. Thus, it makes sense to require that the determination of the weights in this model be through a process of adaptive selection involving environmental and contextual factors, and the accessing of the agent's memory.

%%%%%%%%%%%%%%%%%%%%%%%%%%%%%%%%%%%%%%%%%%

\section{Conclusions}
As I emphasized earlier, the model assumes that at least some complex systems possess a capacity of choice. That may or may not be a valid assumption. However, it should be clear that human behavior is at least partly guided by the assumption that we \textit{do} have free will. Indeed, most legal systems are built on the basis of that assumption. As O'Connor pointed out, then, the freedom that this assumed capacity of choice makes possible should come in degrees and can vary in time with individual agents~\cite{OConnor:2009aa}. Freedom, in this context, is an action taken by an agent and it is this action that the model presented here aims to capture. Given the increasing importance of artificial intelligence and autonomous systems, it also seems worthwhile to compare such actionable freedom across systems. Could a machine make choices as freely as a human? Might there be a way to measure any difference?

In that sense, this model is one of behavior. It considers a system as being contained within an environment that helps constrain that behavior. To a certain extent, there is a similarity here to the cause and effect repertoires of integrated information theory (IIT)~\cite{Oizumi:2014aa} and, indeed, the relationship is part of a larger research project that is in the early stages. Additional work is required to determine the form of the constant of proportionality in Equation~\eqref{freechoice} and determining the exact form for the time function $T$. I suspect both may need to be determined through experiment. In addition, the model presented here employs simple Gaussian distributions, but it would be interesting to see if Fermi--Dirac, Bose-Einstein, or other such distributions give substantially different predictions.

Clearly, the problem of free will is deeply tied to the problem of consciousness. As such, any model will include some of the same baggage as models of consciousness including, for instance, how to deal with qualia which appear to play a role in weighting the choices. More likely than not, the model will need considerable revision and alteration in order to truly capture what we think of when we think of free will, but it at least provides a basic framework for thinking about the problem.

\begin{acknowledgements}
This research was funded by FQXi and the Fetzer Franklin Fund, Grant Nos. FQXi-MGB-1807 and FQXi-RFP-IPW-1911. I would like to thank George Ellis for his persistence in getting me to understand downward causation. I would also like to thank Carlo Rovelli, Jan Walleczek, Lorenzo Maccone, Dominik \v{S}afr\'{a}nek, Nathan Argaman, Vitaly Vanchurin, Nate Durham, and members of the 2019 Modelling Consciousness Workshop in Dorfgastein, Austria for helpful discussions. Additionally, I would like to thank the anonymous reviewers for constructive feedback. Finally, I would like to thank Robert Prentner for encouraging this line of thinking and for detailed comments on the article. 
\end{acknowledgements}

\bibliographystyle{plain}
\bibliography{ModelFreeWill.bib}

\begin{thebibliography}{10}

\bibitem{Bell:2004aa}
John Bell.
\newblock Free variables and local causality.
\newblock In {\em Speakable and {U}nspeakable in {Q}uantum {M}echanics}, pages
  100--103. Cambridge University Press, Cambridge, second edition, 2004.

\bibitem{Berto:2017aa}
Francesco Berto and Jacopo Tagliabue.
\newblock Cellular {A}utomata.
\newblock In Edward~N. Zalta, editor, {\em The {S}tanford {E}ncyclopedia of
  {P}hilosophy}. Metaphysics Research Lab, Stanford University, {F}all 2017
  edition, 2017.

\bibitem{Conway:2006aa}
John Conway and Simon Kochen.
\newblock The {F}ree {W}ill {T}heorem.
\newblock {\em Foundations of Physics}, 36(10):1441--1473, 2006.

\bibitem{Conway:2009aa}
John Conway and Simon Kochen.
\newblock The strong free will theorem.
\newblock {\em Notices of the American Mathematical Society}, 56(2):226--232,
  2009.

\bibitem{DAriano:2014aa}
Giacomo~Mauro D'Ariano, Franco Manessi, and Paolo Perinotti.
\newblock Determinism without causality.
\newblock {\em Physics Scripta}, 2014(T163):014013, 2014.

\bibitem{Durham:2018aa}
Ian~T. Durham.
\newblock God's {D}ice and {E}instein's {S}olids.
\newblock In Anthony Aguirre, Brendan Foster, and Zeeya Merali, editors, {\em
  Wandering Towards a Goal: How Can Mindless Mathematical Laws Give Rise to
  Aims and Intention?}, Frontiers Collection, pages 133--144. Springer, Berlin,
  2018.

\bibitem{Ellis:2016aa}
George~F.R. Ellis.
\newblock {\em How Can Physics Underlie the Mind?}
\newblock Springer, Berlin, 2016.

\bibitem{Fine:2017aa}
Arthur Fine.
\newblock The {E}instein-{P}odolsky-{R}osen {A}rgument in {Q}uantum {T}heory.
\newblock In Edward~N. Zalta, editor, {\em The {S}tanford {E}ncyclopedia of
  {P}hilosophy}. Metaphysics Research Lab, Stanford University, winter, 2017
  edition, 2017.

\bibitem{Hadley:2018aa}
Mark Hadley.
\newblock A {D}eterministic {M}odel of the {F}ree {W}ill {P}henomenon.
\newblock {\em Journal of Consciousness Exploration \& Research}, 9(1):1--19,
  2018.

\bibitem{Holland:1992aa}
J.H. Holland.
\newblock {\em Adaptation in Natural and Artificial Systems}.
\newblock MIT Press, Cambridge, 1992.

\bibitem{Juarrero:2009aa}
Alicia Juarrero.
\newblock Top-{D}own {C}ausation and {A}utonomy in {C}omplex {S}ystems.
\newblock In Nancy Murphy, George~F.R. Ellis, and Timothy O'Connor, editors,
  {\em Downward {C}ausation and the {N}eurobiology of {F}ree {W}ill},
  Understanding Complex Systems, pages 83--102. Springer, Berlin, 2009.

\bibitem{:2002aa}
R.~Kane, editor.
\newblock {\em The Oxford Handbook of Free Will}.
\newblock Oxford University Press, Oxford, 2002.

\bibitem{Libet:1985aa}
Benjamin Libet.
\newblock Unconscious cerebral initiative and the role of conscious will in
  voluntary action.
\newblock {\em The Behavioral and Brain Sciences}, 8(4):529--566, 1985.

\bibitem{Mahalanobis:1936aa}
Prasanta~Chandra Mahalanobis.
\newblock On the generlised distance in statistics.
\newblock {\em Proceedings of the National Institute of Sciences of India},
  2(1):49--55, 1936.

\bibitem{Maoz:2014aa}
Uri Maoz, Liad Mudrik, Ram Rivlin, Ian Ross, Adam Mamelak, and Gideon Yaffe.
\newblock On {R}eporting the {O}nset of the {I}ntention to {M}ove.
\newblock In A.R. Mele, editor, {\em Surrounding free will: Philosophy,
  psychology, and neuroscience}, pages 184--202. Oxford University Press,
  Oxford, 2014.

\bibitem{:2009aa}
Nancy Murphy, George~F.R. Ellis, and Timothy O'Connor, editors.
\newblock {\em Downward {C}ausation and the {N}eurobiology of {F}ree {W}ill}.
\newblock Springer, Berlin, 2009.

\bibitem{Newsome:2009aa}
William~T. Newsome.
\newblock Human {F}reedom and ``{E}mergence''.
\newblock In Nancy Murphy, George~F.R. Ellis, and Timothy O'Connor, editors,
  {\em Downward {C}ausation and the {N}eurobiology of {F}ree {W}ill},
  Understanding Complex Systems, pages 53--62. Springer, Berlin, 2009.

\bibitem{Nozick:1995aa}
Robert Nozick.
\newblock Choice and {I}ndeterminism.
\newblock In Timothy O'Connor, editor, {\em Agents, {C}auses, and {E}vents:
  {E}ssays on {I}ndeterminism and {F}ree {W}ill}, pages 101--114. Oxford
  University Press, Oxford, 1995.

\bibitem{OConnor:2009aa}
Timothy O'Connor.
\newblock Conscious {W}illing and the {E}merging {S}ciences of {B}rain and
  {B}ehavior.
\newblock In Nancy Murphy, George~F.R. Ellis, and Timothy O'Connor, editors,
  {\em Downward {C}ausation and the {N}eurobiology of {F}ree {W}ill},
  Understanding Complex Systems, chapter Ten, pages 173--186. Springer, Berlin,
  2009.

\bibitem{Oizumi:2014aa}
Masafumi Oizumi, Larissa Albantakis, and Giulio Tononi.
\newblock From the {P}henomenology to the {M}echanisms of {C}onsciousness:
  {I}ntegreated {I}nformation {T}heory 3.0.
\newblock {\em PLoS Computational Biology}, 10(5):e1003588, 2014.

\bibitem{Ray:2005aa}
Surajit Ray and Bruce~G. Lindsay.
\newblock The topography of multivariate normal mixtures.
\newblock {\em Annals of Statistics}, 33(5):2042--2065, 2005.

\bibitem{Shepherd:2012aa}
Joshua Shepherd.
\newblock Free will and consciousness: {E}xperimental studies.
\newblock {\em Consciousness and Cognition}, 21(2):915--927, 2012.

\bibitem{Soon:2008aa}
Chun~Siong Soon, Marcel Brass, Hans-Jochen Heinze, and John-Dylan Haynes.
\newblock Unconscious determinants of free decisions in the human brain.
\newblock {\em Nature Neuroscience}, 11(5):543--545, 2008.

\end{thebibliography}

\end{document}